\documentclass[a4paper, USenglish, autoref, final]{lipics-v2021}

\usepackage{hyperref}
\usepackage[colorinlistoftodos,prependcaption,textsize=small]{todonotes}
\usepackage{booktabs} 
\usepackage{tabularx} 
\usepackage{ragged2e} 
\usepackage[framemethod=TikZ]{mdframed} 
\usepackage{xcolor}
\usepackage{graphicx}
\usepackage{subcaption}
\usepackage{csquotes}
\usepackage{wrapfig}

\nolinenumbers

\mdfdefinestyle{contextbox}{
    linecolor=gray!40,
    linewidth=1pt,
    roundcorner=5pt,
    backgroundcolor=gray!5,
    innertopmargin=10pt,
    innerbottommargin=10pt,
    innerleftmargin=10pt,
    innerrightmargin=10pt,
    frametitleaboveskip=5pt
}

\title{Trust-Calibrated Code Review: A Participatory Design Study of Review Workflows for LLM-Generated Multi-File Changes}
\titlerunning{Trust-Calibrated Code Review for LLM-Generated Multi-File Changes}
\author{Lo {Gullstrand Heander}}{Lund University, Lund, Sweden}{lo.heander@cs.lth.se}{https://orcid.org/0000-0002-0695-4580}{}

\author{Agnia Sergeyuk}{JetBrains Research, Belgrade, Serbia}{agnia.sergeyuk@jetbrains.com}{https://orcid.org/0009-0001-1495-9824}{}

\author{Ilya Zakharov}{JetBrains Research, Belgrade, Serbia}{ilia.zaharov@jetbrains.com}{https://orcid.org/0000-0001-7207-9641}{}

\author{Emma Söderberg}{Lund University, Lund, Sweden}{emma.soderberg@cs.lth.se}{https://orcid.org/0000-0001-7966-4560}{}

\author{Nikita Mukhortov}{JetBrains, Amsterdam, Netherlands}{nikita.mukhortov@jetbrains.com}{https://orcid.org/0009-0008-3882-8969}{}

\authorrunning{Gullstrand Heander et al.}

\Copyright{Lo Gullstrand Heander, Agnia Sergeyuk, Ilya Zakharov, and Emma Söderberg}

\funding{The Swedish strategic research environment ELLIIT, the Wallenberg AI, Autonomous Systems and Software Program (WASP) funded by the Knut and Alice Wallenberg Foundation, and JetBrains Research}
\acknowledgements{The authors would like to express their sincere gratitude to the Data Sanity and JS Belgrade communities for their invaluable assistance and support in recruiting participants for this study. We also thank the participants in the workshops and pilot workshops for their knowledge, expertise, and engagement.}

\newcommand{\RQ}[1]{\textbf{RQ\textsubscript{#1}}}
\newcommand{\Section}[1]{Section~\ref{#1}}
\newcommand{\Figure}[1]{Figure~\ref{#1}}
\newcommand{\Table}[1]{Table~\ref{#1}}



\hideLIPIcs

\begin{document}

\maketitle

\begin{abstract}
\noindent
\textbf{Background:} Developers increasingly review multi-file code changes generated by LLM-based agents, yet no validated end-to-end workflow or IDE tooling design exists for this scenario.

\noindent
\textbf{Aims:} We investigate (RQ1) the challenges developers face when reviewing LLM-generated multi-file changes and (RQ2) how developers envision effective workflows for this task.

\noindent
\textbf{Method:} In collaboration with JetBrains, we conducted a participatory design study structured using the double-diamond design process with Discover, Define, Develop, and Deliver phases. Industry practitioners participated in the Discover phase (N=17); seven of these returned for the Develop phase. The Define phase was an author-led synthesis. The Deliver phase produced a conceptual design and a high-fidelity semi-interactive prototype evaluated through a follow-up survey with N=43 practitioners.

\noindent
\textbf{Results:} Participants identified trust-calibration as the central challenge. The study yielded a three-level review workflow (overview, file-analysis, code snippet review) supported by seven design constructs (chunk, risk-per-line, risk-per-file, judge, walk-through, zooming in/out, and security cage). In the validation survey, all three workflow levels scored above the neutral midpoint (means $3.50$--$3.91$ on a five-point scale). Of the respondents, 63\% expected reduced overall review effort, and 52\% reduced trust-assessment effort, relative to their current tools. These findings suggest that the design constructs indicate a positive direction for future tool development.

\noindent
\textbf{Conclusions:} Reviewing LLM-generated multi-file changes is a trust-calibration problem rather than a diffing problem. The three-level workflow and the seven constructs we report give tool designers a conceptual framework for building AI-ready code review tools that surface risk and confidence signals at the granularity at which developers allocate attention.

\noindent
\textbf{Data Availability:} \url{https://doi.org/10.5281/zenodo.20124352}
\end{abstract}

\section{Introduction}
\label{sec:introduction}

Development teams that adopt LLM-based code generation are expecting substantial productivity gains, but reviewing the large amounts of code generated by LLMs is a new challenge that can easily become a bottleneck in the software delivery process. Change sets are larger, human reviewers cannot ask the author about reasoning or intent, and the LLM signals uniform confidence across high- and low-risk code segments~\cite{trust_aigen_tse2026}. LLM-generated code seems harder, not easier, to review. This potentially leads to surface-level reviews, declining code quality, and bugs in production. In collaboration with JetBrains, a leading software tools provider, we conducted a participatory design study to develop tools that support code review of LLM-generated code effectively. 

While substantial work addresses automated code review (where LLMs review code), this work examines augmentation: designing interfaces and workflows that support and improve human judgment when reviewing LLM-generated code. Consider a concrete scenario: Junie~\footnote{Junie, the AI coding agent by JetBrains, \url{https://www.jetbrains.com/junie/}} (an LLM-based coding agent) submits a change touching 10 files in your repository. As the reviewer, how do you assess whether this change is correct, maintainable, and aligned with your architecture? This scenario, reviewing substantial multi-file changes from LLM agents, is the focus of this work. We investigate what features and workflows developers need to effectively review LLM-generated code while preserving the benefits of modern code review~\cite{bacchelli2013expectations}. Our aim is to design tooling that augments rather than replaces human reviewers~\cite{heander_support_2025}.

Based on existing literature, we consider code review (CR) as a process with four stages: Context building and orientation, Inspection of the change, Discussion and iteration (\enquote{shepherding the change}), and Decision to merge~\cite{gonccalves2022explicit,gullstrand_heander_code_2026}. In code review of LLM-generated code, the human reviewer is often also the co-author, possessing a deep knowledge of the codebase, the architecture, and the rationale of the code. Because the co-author is also the reviewer, the orientation phase collapses and the critical need for support emerges during the iterative analytical phase. Unlike conventional code review, where acceptance decisions are typically final, LLM-generated code review involves iterative refinement cycles. Reviewers may accept suggestions conditionally, which triggers further LLM iterations and makes the decision process more dynamic and provisional rather than binary. This iterative nature shifts the reviewer's role toward that of a collaborator and auditor, necessitating new interface paradigms and workflow support tools.

\textbf{\RQ{1}} What are the challenges for developers when reviewing LLM-generated multi-file changes in their IDE?

\textbf{\RQ{2}} How do developers envision effective workflows for reviewing LLM-generated multi-file changes?

Through a participatory design study of these research questions we report four contributions: (1) an empirical characterization of trust-calibration as the dominant reviewer challenge; (2) a three-level IDE workflow co-designed with seventeen practitioners; (3) a conceptual framework of seven design constructs to support reasoning about the design of AI-ready code review tools; and (4) a high-fidelity prototype validated through a survey of forty-three professional developers.

\section{Background and Related Work}
\label{sec:background}

Modern code review is an informal, lightweight, and asynchronous process in which one or more developers inspect a peer's proposed change before it is merged into the shared codebase~\cite{rigby2013convergent, sadowski2018modern}. Although defect detection remains a stated goal, practitioners and researchers consistently report that the value of review is broader: knowledge transfer, shared ownership, design feedback, and informal mentoring~\cite{bacchelli2013expectations, sadowski2018modern}. Recent cognitive accounts characterize the activity as a sequence of decision-making episodes that move through orientation, inspection, shepherding, and a final integration decision~\cite{gullstrand_heander_code_2026, gonccalves2022explicit}. The activity is, in other words, a human-centered exercise in which a reviewer reconstructs a peer's intent and weighs it against project context.

When code is LLM generated, several assumptions on which this exercise rests no longer hold. The LLM cannot be interrogated about its intent; the rationale that a peer reviewer would normally elicit through discussion is absent or, at best, available as a chain-of-thought transcript that is itself subject to confabulation~\cite{trust_aigen_tse2026}. Also, while the strengths and weaknesses of a human author can be learned, the competence profile of an LLM is stochastic and uneven. The change set is typically larger and spans more files than a human-authored commit~\cite{zhong2026human}, yet confidence is presented uniformly across heterogeneous segments, leaving reviewers without a means of allocating attention proportional to risk. LLM tooling additionally introduces an action-space concern that has no analog in peer review: an insufficiently sandboxed LLM-tool can modify files, install dependencies, or execute commands as a side effect of the proposed change. The result is a review task in which the established interpersonal benefits are at risk of being lost while new, LLM-specific failure modes appear. The cost falls directly on practitioners; review effort is distributed without regard to risk, and acceptance decisions are made without visibility into LLM reasoning.

Prior solution attempts can be grouped into three categories. Cognitive-load research on human-authored review has produced two-phase reading strategies and chunking guidelines that scale review effort to working-memory limits~\cite{gonccalves2025code, badampudi2023modern}. A second category investigates AI-as-reviewer systems that automatically comment on human-authored changes~\cite{vijayvergiya2024ai}; these adopt an augmentation stance that is orthogonal to the present work, in which the AI is the author and the human the reviewer. A third category, the \emph{Code Review as Decision-Making} (\textbf{CRDM}) cognitive model, positions review as an orientation phase followed by an iterative analytical phase~\cite{gullstrand_heander_code_2026} and provides adjacent prior work motivating focused study of the analytical phase in isolation; we return to its applicability in \Section{sec:discussion}. To our knowledge, this is the first artifact-producing study to elicit reviewer needs for this scenario through participatory design rather than feature speculation, and to validate the resulting workflow with a separate practitioner sample.

\section{Method}
\label{sec:method}

The study design is based on pragmatist epistemology with participatory design values~\cite{10.7551/mitpress/12255.001.0001}, seeking to model the needs of the community through its active participation in research and evaluation. We adopted the Double Diamond design framework \cite{designcouncil2023} to address our research questions. This process supports the transition from problem identification to solution delivery through four distinct phases: \textbf{Discover}, \textbf{Define}, \textbf{Develop}, and \textbf{Deliver}. This framework is particularly well-suited for a participatory design approach~\cite{spinuzzi2005methodology}, as it allows researchers to employ the expertise of the professional community to address complex outstanding issues. Since external participants' time is valuable, and committing to four multi-hour workshops is daunting for most busy software professionals, we prioritized involving them in the expansive Discover and Develop phases. For the contractive Define and Deliver phases, we held workshops internally in the research group. 

\textbf{Workshop~I (Discover)} and \textbf{Workshop~III (Develop)} followed pre-defined protocols that were, in both cases, first tested and refined in pilot workshops held approximately one week before each workshop. The participants in the pilot workshops were colleagues of the authors with a high degree of software and research experience, enabling both full participation and valuable feedback on the workshop protocol. In \textbf{Workshop~I}, participants (professional developers) identified specific challenges inherent in reviewing LLM-generated code and evaluated a set of literature-derived solutions for effective code review by ranking them according to importance and feasibility. For complete workshop protocols, see \Section{sec:data_availability}.

\textbf{Workshop~II (Define)} used a more exploratory format. As study authors, we each familiarized ourselves with the findings of Workshop~I and structured them into thematic clusters. From the themes and the study context, we derived a prioritized list of problems with opportunity for intervention, which were described in more detail and used as input for Workshop~III. 

In \textbf{Workshop~III}, developers used these prioritized problems to co-design review tools through low-fidelity prototyping using pre-drawn IDE mockups and tactile craft materials. As the last exercise in the workshop, they arranged their design sketches into a multi-level workflow for end-to-end code review of LLM-generated code.

For \textbf{Workshop~IV (Deliver)}, we created a conceptual design~\cite{johnson2002conceptual} grounded in a persona and use cases derived from our observations and notes from the two participant workshops. The workshop outputs, the persona, and the use cases were used to systematically produce the conceptual design, which in turn formed the foundation for a high-fidelity semi-interactive prototype. The visual design and user experience of the prototype were developed by author 5, a professional industry designer.

To validate the prototype, we distributed a survey to participants in previous workshops and pilots, and to software professionals via social media. For the full survey questionnaire, see \Section{sec:data_availability}. The survey presented an animated walk-through of the prototype, demonstrating each level of the three-level workflow and the seven design constructs. After the animation, participants rated their agreement with a series of statements on a five-point Likert scale (1 = Strongly disagree, 5 = Strongly agree). Three further Likert statements and two comparative-effort questions allowed participants to rate the overall impression. Free-text questions in each section collected comments on the prototype and suggestions for improvements.

\subsection{Participants}

To ensure the quality and relevance of the co-design process, we employed a rigorous screening process based on three primary criteria: professional experience, AI tool proficiency, and specific experience in reviewing LLM-generated code. Only developers with active roles in software engineering were considered. Participants were also required to have used AI coding assistants (e.g., GitHub Copilot, Cursor, or ChatGPT) for at least three months, with a usage frequency of at least once per week. A critical requirement was experience in reviewing multi-file LLM-generated changes (3 or more files). 
We initially recruited a cohort of professional developers ($N=17$) to participate in the Discover phase. For the Develop phase, 7 of the original 17 participants returned to continue the co-design process. 

\begin{table*}[hbt]
\centering
\caption{Detailed Profiles of Study Participants ($N=17$).}
\label{tab:participants}
\resizebox{\textwidth}{!}{%
\begin{tabular}{|l|l|l|l|l|l|l|l|}
\hline
\textbf{ID} & \textbf{Job Role} & \textbf{Coding Exp.} & \textbf{AI Rev. Exp.} & \textbf{Co. Size} & \textbf{Primary Product Domains} & \textbf{Primary IDE} & \textbf{WSIII} \\ \hline
P1 & SW Eng. & 6--10y & 1--2y & 1,001--5,000 & Cloud, Web Services, Business Apps & IntelliJ & Y \\ \hline
P2 & SW Eng. & 1--2y & < 6mo & Just me & Dev Tools, Rendering, Websites & Cursor & N\\ \hline
P3 & SW Eng. & 6--10y & < 6mo & 2--10 & AR/VR, Mini-apps & Android Studio & Y\\ \hline
P4 & SW Eng. & 6--10y & 1--2y & 51--500 & Dev Tools & PyCharm & N\\ \hline
P5 & SW Eng. & 11--15y & 1--2y & 11--50 & Cloud, Mini-apps, Websites, Business Apps & Cursor & Y\\ \hline
P6 & Data Scientist & 6--10y & 6--12mo & 5,000+ & Finance, IT Infra, Analytical SW & VS Code & N\\ \hline
P7 & SW Eng. & 11--15y & 6--12mo & 1,001--5,000 & Cloud, Finance, Business Apps & PyCharm & N\\ \hline
P8 & SW Eng. & 6--10y & 1--2y & 501--1,000 & Entertainment, Web Services, Business Apps & Goland & N\\ \hline
P9 & DevOps Eng. & 6--10y & 6--12mo & 11--50 & Cloud, Hardware, IT Infra & Cursor & Y\\ \hline
P10 & AI Researcher & 1--2y & 6--12mo & 51--500 & IT Infrastructure, Websites & PyCharm & Y\\ \hline
P11 & SW Eng. & 6--10y & 6--12mo & 51--500 & Security, Websites, Business Apps & PyCharm & N\\ \hline
P12 & AI Researcher & 6--10y & 1--2y & 501--1,000 & Security & VS Code & N\\ \hline
P13 & AI Researcher & 11--15y & 1--2y & 5,000+ & IT Infrastructure & PyCharm & N\\ \hline
P14 & AI Researcher & 11--15y & 6--12mo & 2--10 & Business Apps, Analytical SW, AI Apps & Cursor & Y\\ \hline
P15 & SW Eng. & 1--2y & 1--2y & 11--50 & Business Applications & VS Code & Y\\ \hline
P16 & Physics Res. & 11--15y & 6--10y* & 51--500 & (No specific product) & PyCharm & N\\ \hline
P17 & SW Eng. & 16+y & 1--2y & 51--500 & Bioinformatic Pipelines & RustRover & N\\ \hline
\end{tabular}%
}
\begin{flushleft}
\scriptsize \textit{Note: SW Eng. = Software Engineer; AI Rev. Exp. = Years using AI tools for development/review; Co. Size = Total employees.
WSIII = Participated in Workshop III (all participated in Workshop I). Participants were recruited based on their experience with code review of multi-file LLM-generated code.}
\end{flushleft}
\end{table*}

The participants represented a high level of professional expertise and diverse technical backgrounds (see \Table{tab:participants}). The cohort was notably senior; over 70\% of participants possessed more than 6 years of professional experience, with several exceeding 11 and 16 years in the industry. The majority of participants identified as Software Developers/Engineers, supplemented by AI/ML Engineers and Team Leads. The majority of the participants were using tools such as ChatGPT, Cursor, and GitHub Copilot on a daily basis. Most had been integrating these tools into their workflow for over six months, with several being early adopters (2+ years). All participants were active in code review, with most performing reviews daily or several times per week. Ninety percent of the selected participants confirmed they had reviewed LLM-generated changes involving three or more files \enquote{multiple times}, providing the necessary depth of knowledge for the study.

For the validation survey, we received a total of 97 responses. However, 54 of these responses were flagged as LLM-generated based on time to complete the survey, similarity to other responses, referrer link, and evident prompt leaking. This left $N=43$ legitimate respondents, predominantly experienced practitioners: 70\% reported 6 or more years of professional coding
experience. The most common job roles were software developers or engineers (42\%) and AI/ML engineers (19\%). All participants had used AI
tools for software development for at least six months; 74\% for one year or more. AI tool use was intensive: 91\% used AI tools
at least several times per week. Participants were also active code reviewers, with 70\% reviewing code several times per week
or more. The most common case (37\%) was that 31-60\% of the code reviewed was LLM-generated. Survey participants used several different AI coding tools. The most widely used tools were Claude Code (58\%), ChatGPT (51\%), and OpenAI Codex (47\%). Of the respondents, 35\% had participated in one of the previous workshops while the rest had found the survey on social media (LinkedIn). Complete demographics are available in the replication package, see \Section{sec:data_availability}.

\subsection{Data Analysis}

Workshop~I generated 60 post-it notes capturing 64 participant-identified challenges, each annotated with a severity and feasibility rating. All four study authors collectively coded these notes through an iterative consensus-based process. In a series of joint coding sessions, the authors grouped the notes into thematic clusters, discussing each grouping until interpretative convergence was reached. The fourth author had not facilitated Workshops~I or~III and joined the synthesis at the convergent Workshops~II and~IV, providing an outside reading of the divergent-phase data and a form of investigator triangulation~\cite{denzin2017research} embedded in the team composition. The resulting digitized post-it dataset, complete with thematic coding, is included in the replication package, see (\Section{sec:data_availability}).

Workshop~II continued the analysis: the authors used the thematic clusters from Workshop~I to derive a prioritized list of eleven problems warranting intervention, guided by the severity and feasibility ratings participants assigned during Workshop~I. The prioritization was based on the frequency with which challenges appeared across clusters combined with the participant-assigned ratings, ensuring the final problem list was empirically grounded in the Workshop~I data rather than author preference alone.

Workshop~III produced low-fidelity sketches. These artifacts were analyzed collectively by the authors following the same consensus-based approach, to extract recurring design constructs and a common workflow structure. The construct extraction was driven by the data: constructs were included only when they appeared in multiple participant sketches, were voted for by multiple participants, or were highlighted in workshop discussion transcripts.

The validation survey was analyzed by computing mean Likert scores for each of the levels as well as descriptively by plotting the responses centered around the neutral midpoint (\Figure{fig:likert_chart_clean}). While Likert scores are ordinal variables, research shows that computing means give the correct interpretation especially when summing over several questions and respondents~\cite{normanLikertScalesLevels2010}. The \enquote{Overall} section of the survey contains three singular Likert questions as well as two questions using a scale from \enquote{Much more} to \enquote{Much less}. For these questions, we instead analyze and present the percentages of respondents who choose the different alternatives. Free-text answers were analyzed using thematic analysis by author 1 and author 4 independently, with the themes then combined to form a consensus.
\section{Results}
\label{sec:results}

\subsection{Discover Phase}

\begin{wrapfigure}[22]{L}{0.4\textwidth}
    \centering
    \includegraphics[width=0.95\linewidth]{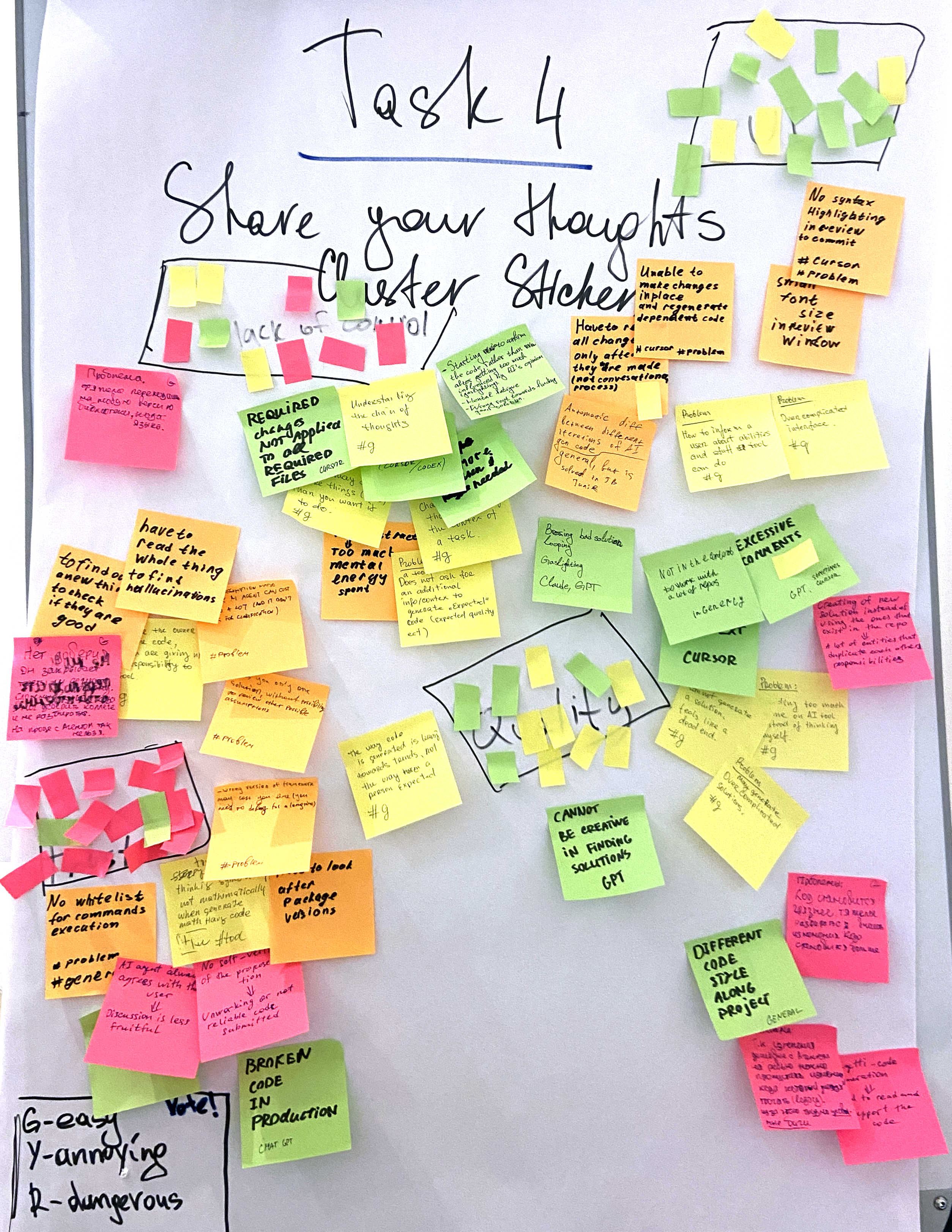}
    \caption{Post-it notes with challenges, grouped into clusters by the participants. \textbf{Takeaway:} The trust cluster received the most high-severity votes.}
    \label{fig:ws1_post_it_notes}
\end{wrapfigure}

In the Discover phase workshop, the participants contributed 64 post-it notes capturing challenges in reviewing LLM-generated code, categorized into four groups: `Trust', `Lack of control', `Quality', and `UI'. The small paper strips in green, yellow, and red (see \Figure{fig:ws1_post_it_notes}) denote votes for the groups as `easy', `annoying', and `dangerous', respectively. Trust dominated on severity, receiving 12 of the 17 red (dangerous) votes cast across all groups; Lack of control attracted the remaining 5 red votes, while Quality and UI received yellow and green votes only. For full per-group vote counts and digitized post-it notes, see \Section{sec:data_availability}. 

\subsection{Define Phase}

The challenges and groups of the discovery phase were further refined in the define phase into 12 topics: `Broken code' (13 occurrences), `Intent misalignment' (8), `Bad UI/UX' (6), `Overcomplexity' (6), `Cognitive load' (6), `Lack of trust' (5), `Lack of transparency/explainability' (5), `Inconsistent style' (4), `Insufficient context window' (4), `Increased tech debt' (3), `Redundancy' (2), and `Overtrust/overreliance' (2). Although `Broken code' and `Intent misalignment' attracted the highest occurrence counts, these challenges are more appropriately addressed during code generation rather than during code review. In addition to `Lack of trust' directly, several of the remaining topics are indirectly linked to trust, such as `Lack of transparency/explainability', `Overtrust/overreliance', and parts of `Cognitive load'. Trust-related topics surface as the most salient for the review stage: the trust cluster alone attracted 12 of the 17 red (dangerous) votes in the workshop. We therefore focus the problem definition on this theme.

\subsubsection{Problem Background}
Trust is complicated in the case of review and integration of LLM-generated code, especially when the agent creates or changes multiple files to solve a more complex issue description or prompt. The problems center on several themes:

\textbf{Explainability.} The agent gives no information about its chain of thought, references and why it chose certain solutions. Different implementation options are not presented, neither the assumptions applied. Makes it hard to judge if the solution is good enough and is matching functional requirements.

\textbf{Transparency.} Different segments in different files could have very different levels of confidence when generated, but are presented as equal. This can force the human reviewer to very carefully check all code, spending a lot of time on the review.

\textbf{Inconsistent attention to detail.} Even if specified in the prompt or context, the agent might get package versions for APIs and libraries wrong (likely because a certain version was most common in its training data). This results in code that looks correct but will fail or behave in unpredictable ways.

\textbf{Insufficient sandboxing.} The agents might be insufficiently sandboxed and able to execute too many commands in the programming environment, sparking worry of files or environment getting compromised.

\textbf{Overconfidence and gaslighting.} The agent will write in a manner similar to a human who is very sure of their solution, but can also reverse direction completely to agree with the human user if challenged. This can confuse the user and erode trust in the solution.

\subsubsection{Problem Definition}
An appropriate level of trust is necessary to complete the review of LLM-generated code in an efficient way and without too much mental fatigue and frustration. This may result in low trust, particularly for complex code, where the agent exhibits low internal confidence (the model’s internal uncertainty about the generated output), or high trust where the code is simple and the probability of correct tokens is high. The problem lies in giving the human reviewer enough information and guiding them to place the correct level of trust in the correct parts of the solution. Achieving this can mean that users experience confidence in the system as a whole, since they trust it to show accurate information about potential issues and code sections that need more careful review. 

The user wants to: A) verify the LLM-proposed solution to avoid merging bugs, security risks, performance bottlenecks, and other issues into the production code base. B) efficiently evaluate very large and complex LLM-generated code changes because reading through all the lines is error-prone, creates mental fatigue, and is time consuming.

\subsection{Develop Phase}

In the Develop Phase workshop, the participants created short solution suggestions written on post-it notes, design sketches for some of these solutions, and workflows describing where some of these design sketches could fit into a code review workflow. The participants proposed a three-level workflow: 1) overview, 2) file-analysis, and 3) code snippet review.

\begin{figure}[ht]
    \centering
    \begin{subfigure}[t]{0.38\textwidth}
    \centering
    \includegraphics[width=\textwidth]{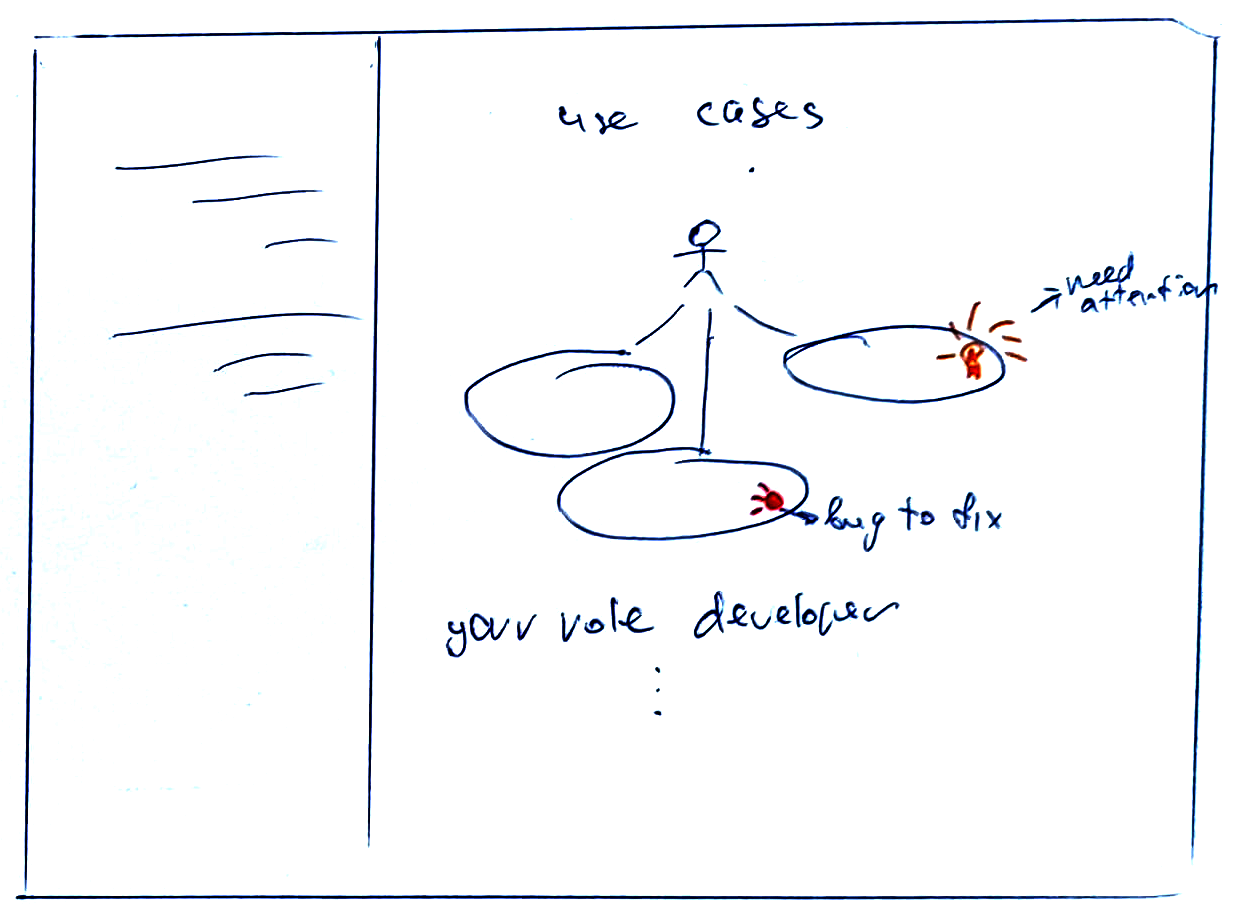}
    \caption{Sketch of architectural diagram highlighting code changes and potential issues.}
    \label{fig:arch_diagram}
    \end{subfigure}
    \hfill
    \begin{subfigure}[t]{0.38\textwidth}
    \centering
    \includegraphics[width=\textwidth]{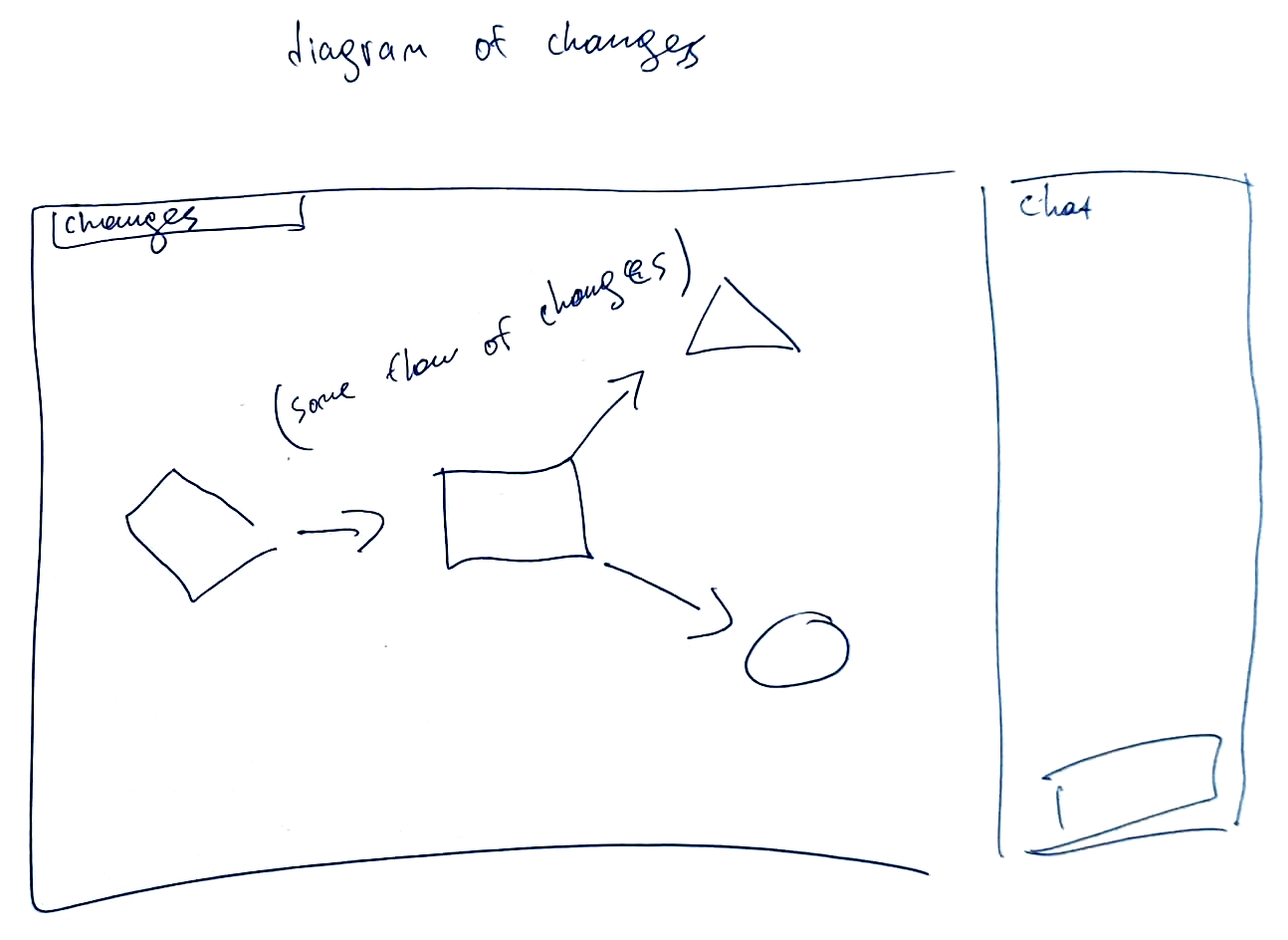}
    \caption{Sketch of data flow diagram.}
    \label{fig:dataflow_diagram}
    \end{subfigure}
    \vspace{1em}
    \begin{subfigure}[t]{0.38\textwidth}
    \centering
    \includegraphics[width=\textwidth]{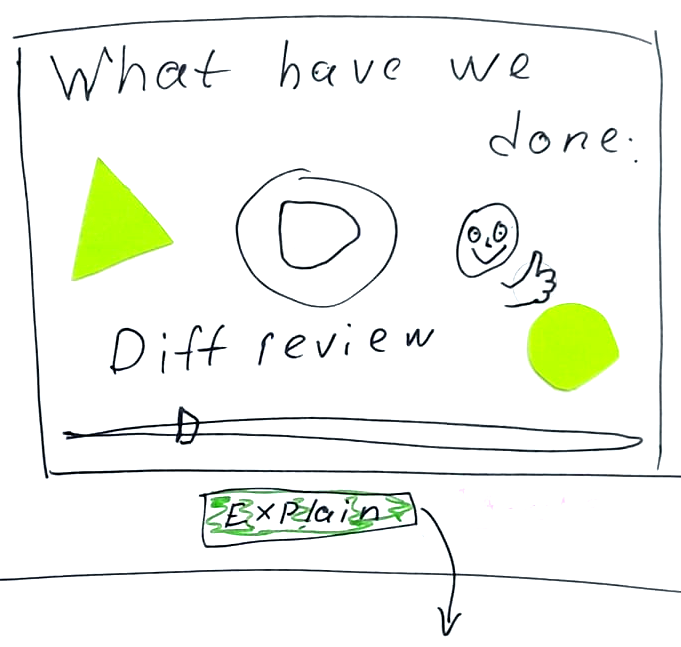}
    \caption{Sketch of an LLM-generated walk-through explaining the rationale and implementation choices behind the proposed change.}
    \label{fig:video_walkthrough}
    \end{subfigure}
    \hfill
    \begin{subfigure}[t]{0.38\textwidth}
    \centering
    \includegraphics[width=\textwidth]{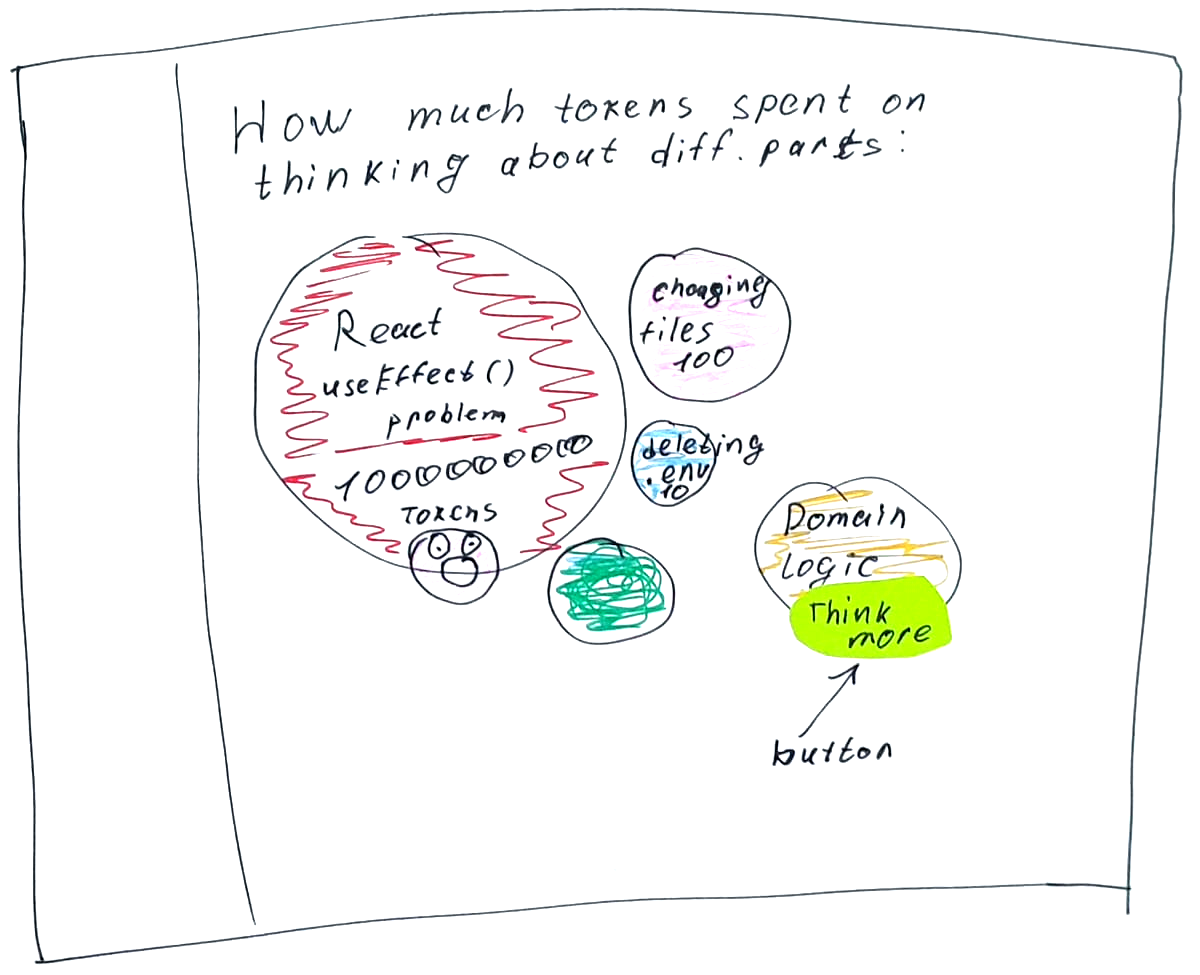}
    \caption{Sketch of token cost diagram.}
    \label{fig:token_costs}
    \end{subfigure}
    \caption{Participant sketches for the \emph{overview} level. \textbf{Takeaway:} Overview level zooms out and provides abstract perspectives on the LLM-generated solution.}
    \label{fig:overview_phase}
\end{figure}

In the \emph{overview} level, the user can choose from several different views (\Figure{fig:overview_phase}) that give higher abstraction-level explanations of the proposed code changes. Two proposed views highlight the changes and potential issues in diagram form, as an architectural diagram (\Figure{fig:arch_diagram}) and a data flow diagram (\Figure{fig:dataflow_diagram}). A third proposal introduces an LLM-generated walkthrough of the code and the reasoning behind it (\Figure{fig:video_walkthrough}). The final view shows tokens spent by the LLM on the different parts of the code and the design, to give the reviewer an indication if some parts have been neglected or overcomplicated (\Figure{fig:token_costs}).

\begin{figure}[ht]
    \centering
    \begin{subfigure}[t]{0.19\textwidth}
    \includegraphics[width=\textwidth]{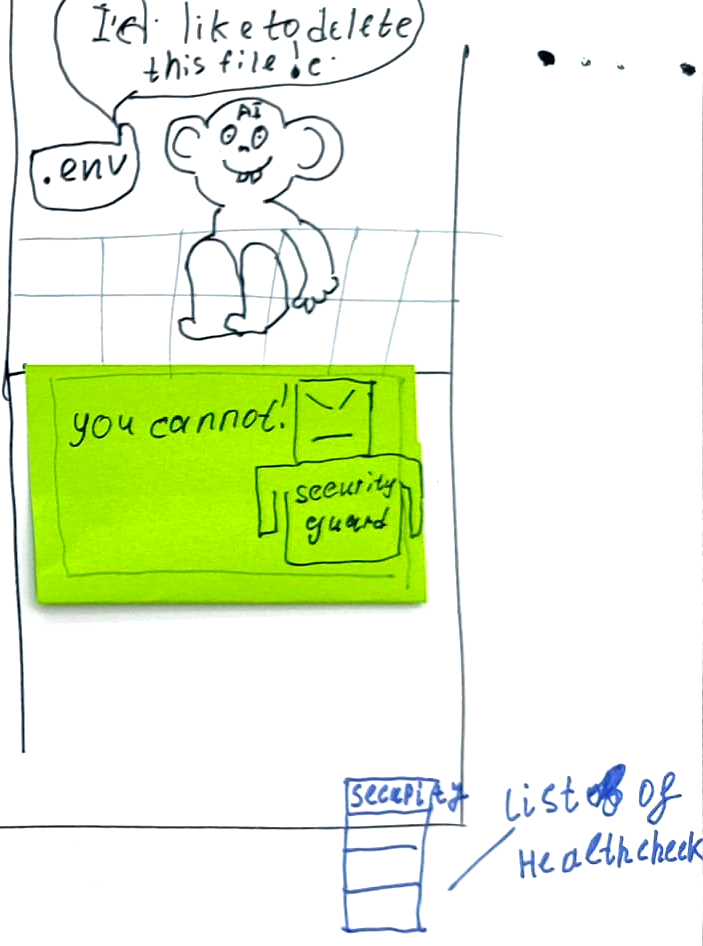}
    \caption{Sketch of the security cage view.}
    \label{fig:cage_view}
    \end{subfigure}
    \hfill
    \begin{subfigure}[t]{0.36\textwidth}
    \includegraphics[width=\textwidth]{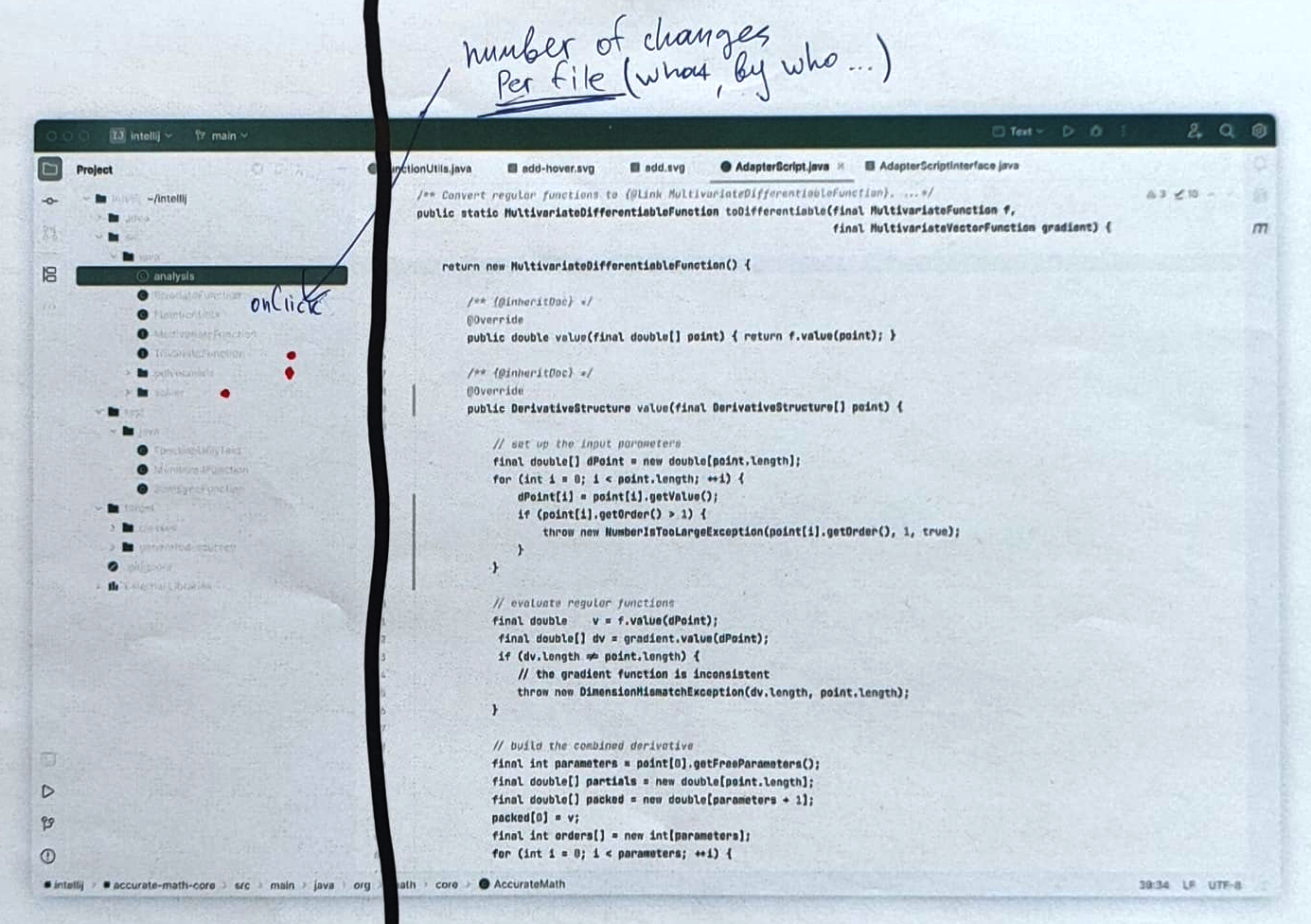}
    \caption{Sketch of the risk analysis view.}
    \label{fig:risk_view}
    \end{subfigure}
    \hfill
    \begin{subfigure}[t]{0.36\textwidth}
    \includegraphics[width=\textwidth]{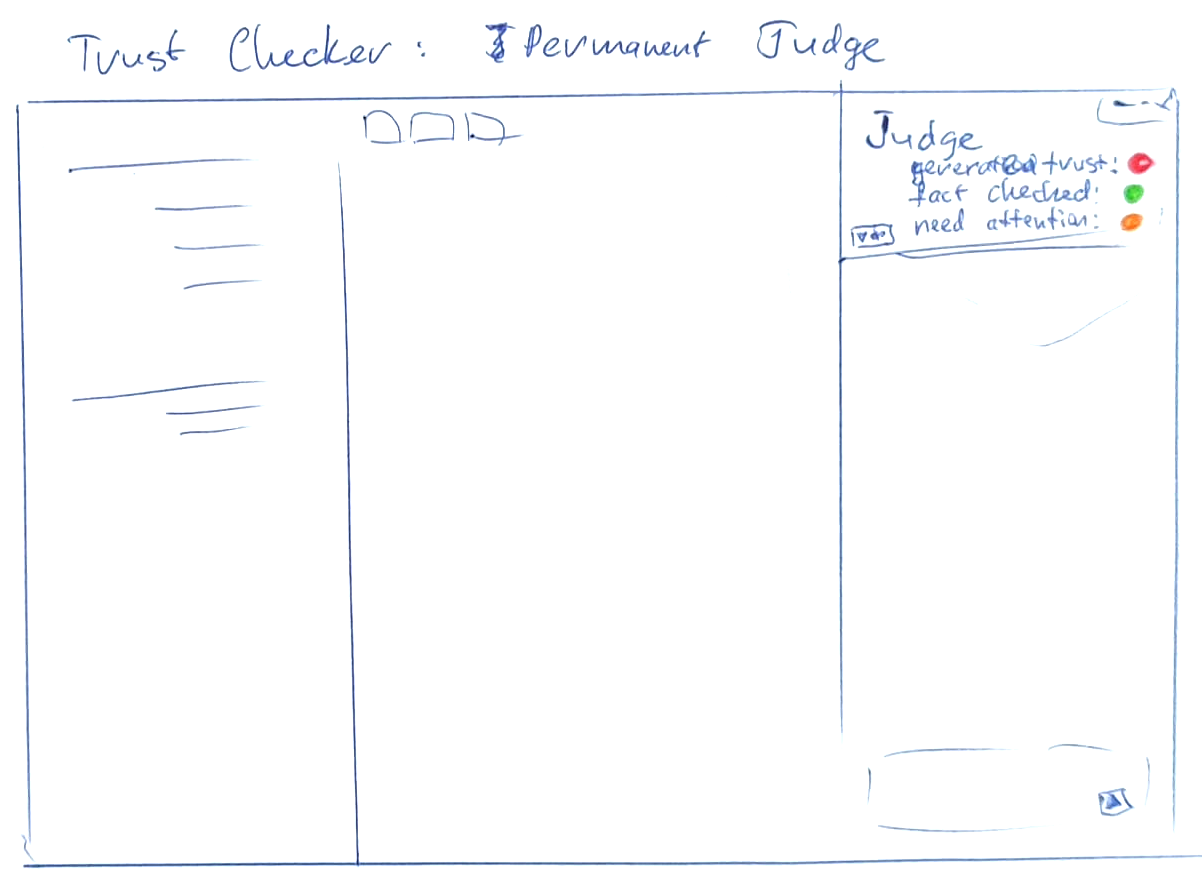}
    \caption{Sketch of the judge view.}
    \label{fig:judge_view}
    \end{subfigure}
    \caption{Participant sketches for the \emph{file-analysis} level. \textbf{Takeaway:} File-analysis level zooms in for whole-file assessment concerning trust and risk calibration.}
    \label{fig:file_level_phase}
\end{figure}

After the reviewer understands the solution proposal at the architectural and conceptual levels, they move to the \emph{file-analysis} level (\Figure{fig:file_level_phase}) where the IDE guides them through assessing one file at a time. Three proposed views are central to this level. The first is a security cage that limits the actions the agent can take in the user's IDE, for increased safety and trust. The cage also provides a safe environment for running health checks on the proposed solution and alerts the reviewer to potential failures and security risks. The second is a file-level and line-level risk analysis based on the frequency of change of files and lines (\Figure{fig:risk_view}). If the file as a whole, or specific lines, have been changed very rarely throughout the project's history, the changes are considered higher risk than if the file and lines are changed very frequently. The third proposed view was named the `Judge' view by the workshop participants (\Figure{fig:judge_view}). This view shows an LLM's assessment of the code, acting as a judge. The judge performs issue detection, fact checking, security verification, and presents areas needing attention by the reviewer as traffic lights.

\begin{wrapfigure}[17]{L}{0.49\textwidth}
    \centering
    \includegraphics[width=\linewidth]{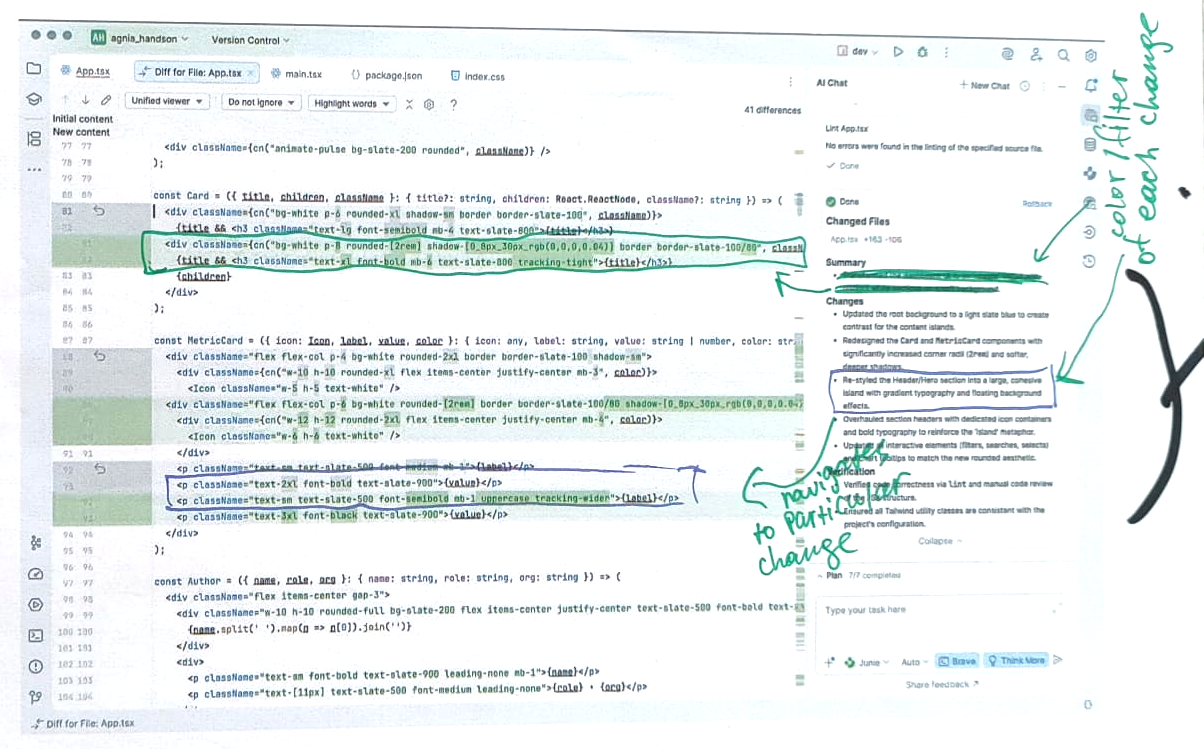}
    \caption{Participant sketch for the \emph{code snippet review} level. \textbf{Takeaway:} Code snippet review level zooms all the way in, and breaks the proposed solution into \emph{chunks}, the smallest logically cohesive code units.}
    \label{fig:chunk_level}
\end{wrapfigure}

Finally, in the \emph{code snippet review} level, a view breaks down the proposed solution into its smallest independent parts: \emph{chunks}. Each chunk is a set of changes across one or more files that belong together. They must be applied as a unit for the code to compile and the test suite to pass. A proposed solution could consist of a single chunk if all changes depend on each other, or of several chunks if the solution has several orthogonal parts that can be applied independently. Breaking the solution into chunks helps the reviewer understand, analyze, and approve it at a more granular level. As seen in \Figure{fig:chunk_level}, each chunk is linked to the corresponding parts of the prompt and the chain-of-thought segments in the LLM response.

\subsection{Conceptual Design}

The sketches and workflows from the Develop phase were refined by the authors into a conceptual design~\cite{johnson2002conceptual}.

\textbf{Concept: Proposed Solution.}
The proposed solution is the whole of all proposed file changes by the LLM with the aim of fulfilling the task described in the latest iteration of the prompt. The user can change it manually or ask the LLM to modify it, before merging it into their working directory.

\textbf{Concept: Risk-per-line.}
Risk per line is the concept that different source code lines in a project carry different risk if they are modified. Some lines might be very important to the logic, performance, or security of the code, while others control more peripheral behavior. 

\textbf{Concept: Risk-per-file.}
Similarly to risk per line, changing a file could have higher or lower potential consequences on the behavior of the system as a whole.

\textbf{Concept: Chunk.}
A chunk is the smallest subset of code changes in the proposed solution that could be applied together without breaking code compilation or test suites.

\textbf{Concept: Token Cost.}
When LLMs generate solutions, their providers often charge the users for the number of `tokens' processed as input and generated as output. Knowing and controlling this cost can help the user balance the complexity, size, and cost of the proposed solution depending on their needs and their budget.

\textbf{Metaphor: Walk-through.}
Just like a colleague taking you by the hand and showing you around the factory floor to explain the processes there, in the metaphor of a walk-through the system gives you a guided tour through the proposed solution offering explanations for the rationale, the reasoning, and the implementation choices.

\textbf{Metaphor: Judge.}
In the metaphor of a Judge, the \emph{proposed solution} is analyzed and criticized by an AI agent. In the sketches from the workshop participants, the verdict from the Judge is presented as traffic lights for potential bugs, fact checks, security verification, etc. Areas of the proposed solution that received a red or yellow traffic light from the AI judge are highlighted for closer manual inspection. Explanations and justifications by the AI judge could guide the human engineer in what to look for.

\textbf{Metaphor: Security Cage.}
In the security cage metaphor, the AI agent's actions are restricted by placing well defined limits on what it is allowed to do and surfacing these restrictions to the reviewer. This mitigates action-space risk and helps the reviewer trust that nothing unexpected was changed in their workspace.

\textbf{Metaphor: Zooming in and out.}
Like a telescoping camera lens directed at the code, zooming in and out allows the user to move through the three-level workflow and zoom between the different levels of abstraction they represent.

\textbf{Metaphor: De-tangling.}
Just like finding and separating the individual ropes in a tangled-up bunch, de-tangling code separates complex code changes into chunks connected by shared logic or rationale. This allows the user to analyze the LLM-generated solution part-by-part. Used to create and show the chunks in the \emph{code snippet review} level.

\textbf{Metaphor: Traffic Light.}
When viewing assessments of proposed solution, the traffic light metaphor gives a quick and visual signal of which parts need immediate attention (red light), where there could be potential issues (yellow), and which parts have a high likelihood of being correct (green). Used to communicate AI judge verdict in \emph{file-analysis} level.

\subsection{High-Fidelity Prototype}

\begin{figure}[hbtp]
    \centering
    \includegraphics[width=0.55\textwidth]{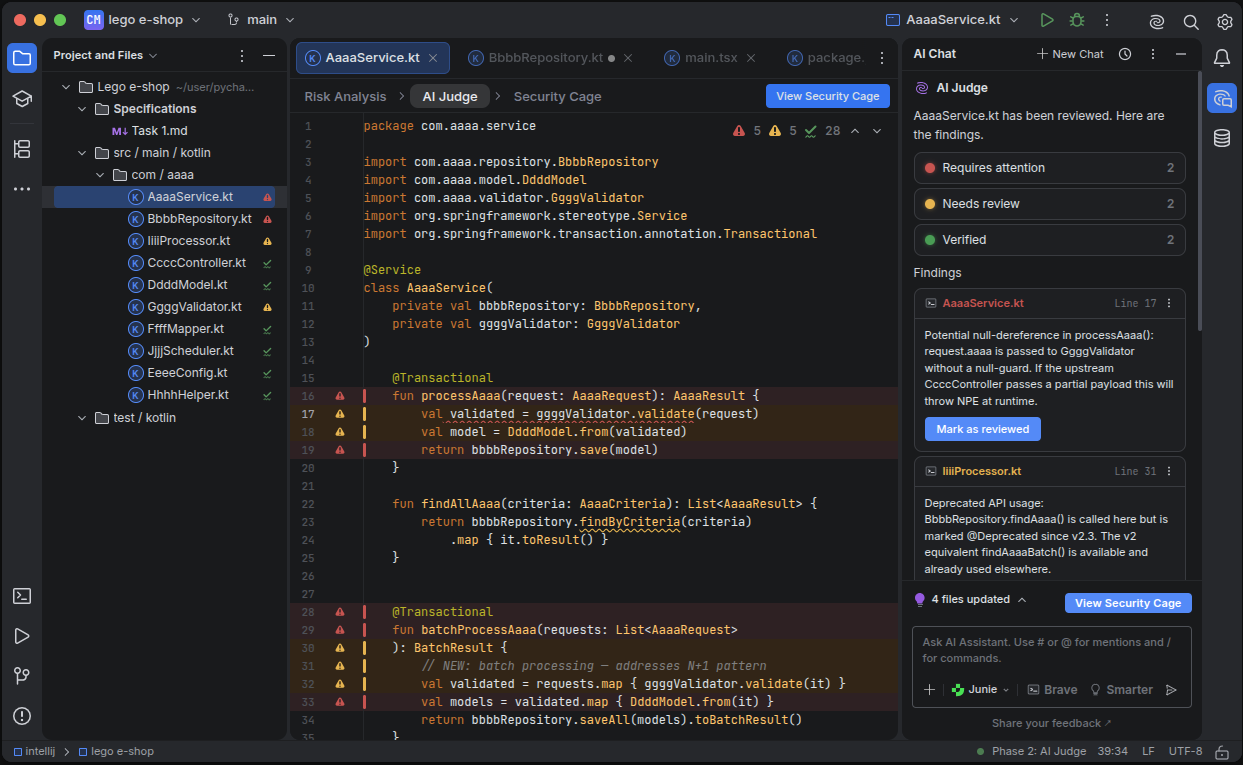}
    \caption{High-fidelity prototype screenshot example: the AI Judge verdict panel from the file-analysis level. \textbf{Takeaway:} The prototype operationalizes all seven constructs and implements the three-level workflow, providing a concrete artifact for survey-based validation.}
    \label{fig:hifi_screenshots}
\end{figure}

The conceptual design and the task scenarios were realized as a high-fidelity, semi-interactive prototype implemented in JavaScript using the React framework. The prototype renders the three-level workflow as an embedded panel within a mock IDE and instantiates each of the seven design constructs, i.e., chunk, risk-per-line, risk-per-file, Judge, walk-through, zooming in/out, and security cage, through dedicated views. Example screenshot in \Figure{fig:hifi_screenshots}. Navigation between levels, file and chunk expansion, Judge verdict toggling, and risk indicator traversal are fully interactive. The underlying LLM assessments, chain-of-thought excerpts, and walk-through narration are pre-rendered and static, avoiding confounds from live model variability while preserving the fidelity needed to elicit informed feedback. 
The animated prototype demonstrations are included in the replication package (\Section{sec:data_availability}).

\subsection{Prototype Validation Survey}

\begin{figure}[ht]
    \centering
    \includegraphics[width=0.9\textwidth]{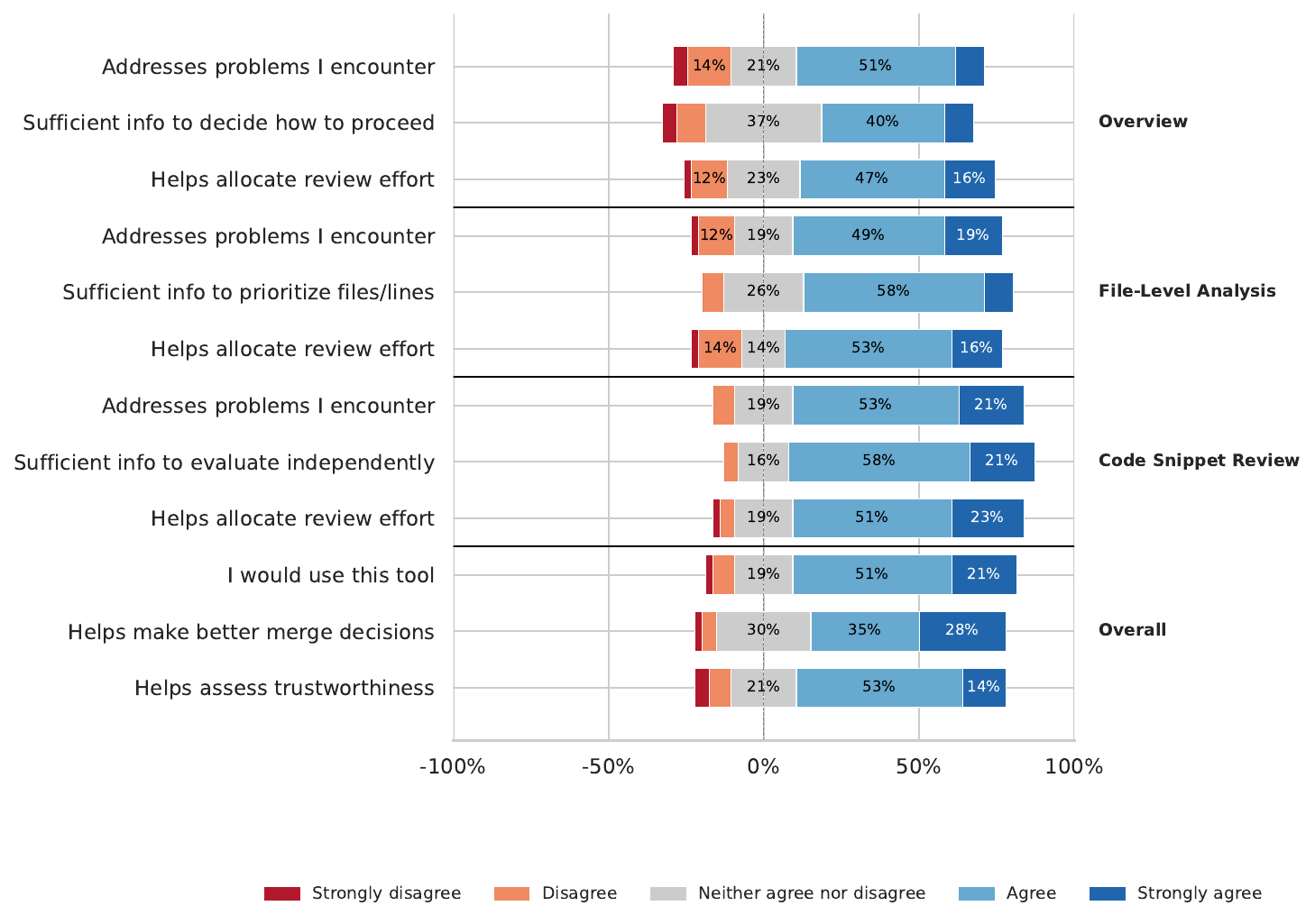}
    \caption{Likert responses for the three levels and overall tool assessment, rated on a five-point scale (1 = Strongly disagree, 5 = Strongly agree). \textbf{Takeaway:} All three levels scored above the neutral midpoint, with the Code Snippet Review level receiving the highest mean rating (3.91), suggesting that chunk-level decomposition resonated most strongly with survey participants.}
    \label{fig:likert_chart_clean}
\end{figure}

\begin{figure}[ht]
    \centering
    \includegraphics[width=0.9\textwidth]{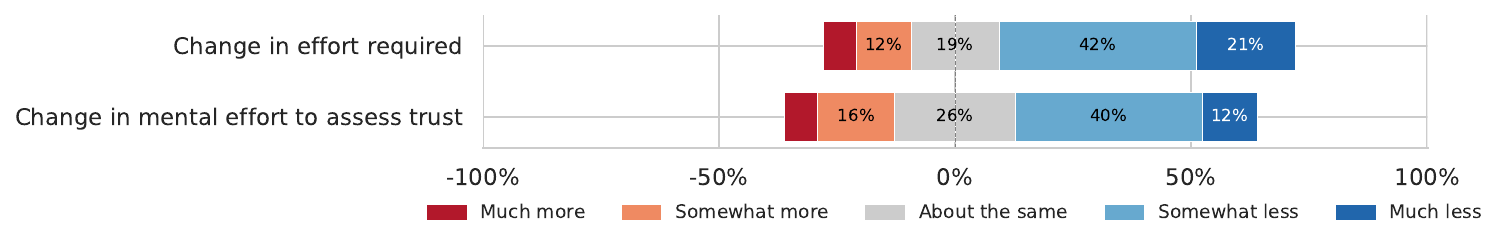}
    \caption{Comparison of perceived effort between the proposed tool and participants' existing code review tools, for overall review and for trust assessment. \textbf{Takeaway:} A majority of respondents estimated that the proposed tool would reduce effort compared to their current tools, both for general review (63\%) and for assessing trustworthiness (52\%).}
    \label{fig:comparative_chart_clean}
\end{figure}

The survey results, summarized in \Figure{fig:likert_chart_clean}, indicate that participants responded positively to all three levels of the proposed workflow. The \emph{Code Snippet Review} level received the highest mean rating (3.91), followed by the \emph{File-Analysis} level (3.69) and the \emph{Overview} level (3.50). All three levels scored above the scale midpoint, suggesting that the workflow structure and its supporting constructs were well-received by the survey participants. That the respondents rated the \emph{Code Snippet Review} level the highest is also reflected in the free-text responses discussed below.

Concerning overall tool assessment, 72\% of respondents agreed or strongly agreed that they would use a tool with these features; 63\% agreed or strongly agreed that the tool would help them make better merge decisions; and 67\% agreed or strongly agreed that the tool would help them assess the trustworthiness of LLM-generated code changes. These findings indicate that guiding reviewers through progressive levels of detail, while surfacing risk and trust signals, aligns with what developers perceive as useful support for the review of LLM-generated multi-file changes.

Regarding comparative effort (\Figure{fig:comparative_chart_clean}), 63\% of respondents estimated that the proposed tool would require less or much less effort than their existing code review tools for the overall review task. For the specific task of assessing trustworthiness, 52\% estimated a reduction in effort. While the trust-assessment figure is more modest, it is noteworthy given that none of the participants' current tools are designed to surface trust signals explicitly. The result suggests that even a prototype demonstration could shift developers' perception of how much effort trust assessment demands.

Free-text responses reinforced and qualified the Likert findings. The \emph{security cage} construct drew the most skepticism: several respondents described its purpose as unclear or redundant, given that sandboxing is typically configured at the agent level rather than surfaced to the reviewer. We retain it in the framework because it is the only construct targeting action-space rather than information-space risk. For the \emph{Code Snippet Review} level, multiple respondents wrote that chunk-level decomposition aids understanding even when chunks cannot be applied independently, reinforcing the high Likert score. Many participants wished for traffic light color-coding of the chunks to convey LLM confidence. Across levels, several respondents preferred free navigation between views rather than a fixed three-level sequence, wanted support for multiple collaborating reviewers, and suggested integrating live test-coverage data into the Judge view. Taken together, the survey results provide initial evidence that the three-level workflow and the seven conceptual design constructs resonate with practitioners' needs. However, as the prototype was semi-interactive and the evaluation was based on a video demonstration rather than hands-on use, these findings should be interpreted as directional rather than conclusive (see \Section{sec:threats}).

\section{Discussion}
\label{sec:discussion}

Prior empirical characterizations of the code review challenge position diff
comprehension, cognitive load, and defect-detection pressure as the primary
obstacles to effective review~\cite{gonccalves2025code, badampudi2023modern}.
These framings are well-motivated for human-authored change sets, where the
reviewer's core difficulty is one of reconstruction: understanding what a
colleague intended, how the change fits the architecture, and whether the
implementation is correct. The present data suggests that, when the author is
an LLM agent, this comprehension framing captures downstream effects rather
than the root cause.

The Discover-phase challenge clusters reveal a more fundamental difficulty.
Reviewers cannot allocate attention proportionate to segment-level risk because
the agent provides no signal about its own confidence or the reasoning behind its
implementation choices. The reviewers fall back to reading every line because any line could be wrong; the rational response to an agent that presents
heterogeneous-quality output with homogeneous confidence. This burden is amplified with agents typically generating large volumes of code, requiring reviewers to inspect substantially more material. Cognitive-load challenges in the data are largely traceable to this opacity rather than to difficulty parsing the code itself.

We therefore propose \emph{trust-calibration}, defined as the capacity to
allocate review effort proportionate to segment-level risk when the author cannot
be interrogated about its confidence or reasoning, as the organizing problem for
reviewing LLM-generated multi-file changes. This framing is consistent with
recent work arguing that trust in AI assistants for software development requires
rethinking when the assistant is the author rather than the
helper~\cite{trust_aigen_tse2026}. The design implication follows directly: tool
designs that address comprehension without addressing trust calibration may
improve efficiency for low-stakes changes while leaving the more consequential
failure mode, systematic misallocation of review effort toward low-risk segments
and away from high-risk ones, unaddressed.

\subsection{A Conceptual Framework for Trust-Signal Exposure}

The co-designed workflow and seven constructs share a common mechanism:
each makes a specific type of epistemic uncertainty visible at a specific
granularity. At change-set granularity, the walk-through, architectural
overview, and token-cost views surface intent and the distribution of
generation effort. At file granularity, risk-per-file and the Judge verdict
prioritize attention before line-level reading begins. At chunk granularity,
risk-per-line and chain-of-thought support decisions around each segment. Line-level details may overwhelm a
reviewer who has not yet obtained architectural orientation, while only file-level and architecture-level information is insufficient for decision-making. The three-level
structure is a proposed answer to that problem.

The structure also has theoretical motivation in prior cognitive models of
review. The \emph{Code Review as Decision-Making} (\textbf{CRDM})
model~\cite{gullstrand_heander_code_2026} describes review as an orientation phase, in
which the reviewer builds a model of what was changed and why, followed by
an iterative analytical phase in which integration decisions are taken. In
human-authored review, orientation draws on interpersonal resources: the
reviewer can ask the author about intent, draw on knowledge of the author's
competence profile, and consult prior discussion threads. When the author is
an LLM, these resources are absent. Orientation does not disappear as a need;
it shifts entirely onto the artifact. The \emph{overview} level responds to
this shift by providing views that substitute for the interpersonal orientation
CRDM implicitly assumes. The \emph{file-level analysis} level inserts a
risk-stratification step into the \emph{assessing change} action in CRDM, made necessary
by the combination of larger change sets and uniform confidence presentation.
The \emph{code snippet review} level recovers the analytical-phase granularity
of prior models, augmented by chunk decomposition and chain-of-thought linkage.
Taken together, the three levels extend the orientation-then-analysis structure
of CRDM by inserting a risk-stratification layer and grounding each analytical
decision in agent-provided reasoning, both of which address conditions that
did not exist when the prior models were developed.

Two constructs merit separate consideration because their standing in the
framework is less settled. The \emph{security cage} is the only construct
targeting action-space rather than information-space risk; survey reception was
mixed (see \Section{sec:results}), which we interpret as reflecting genuine
ambiguity about whether sandboxing is a reviewer concern or an
agent-configuration concern upstream of review. We retain it in the framework
because action-space risk is otherwise unrepresented and because Develop-phase
participants placed it explicitly within the file-analysis level rather than
treating it as infrastructure. The \emph{Judge} construct carries its own
epistemic tension: its verdicts are themselves LLM outputs and therefore subject
to the same opacity that motivates the broader framework. We believe the Judge
is nonetheless useful as an attention-routing device, provided that the interface
communicates that its assessments are probabilistic rather than authoritative,
preserving the reviewer's decision-making role.

Concerning \RQ{1}, the Discover and Define phases converge on trust-calibration,
i.e., the inability to allocate review effort proportionate to segment-level risk
in the absence of transparent confidence signals from the LLM, as the primary
challenge when reviewing LLM-generated multi-file changes; the remaining challenge
topics are interpretable as inputs to or consequences of this problem. Concerning
\RQ{2}, the three-level workflow and the seven constructs constitute a
practitioner-co-designed conceptual framework for IDE tooling that exposes trust
signals at change-set, file, and chunk granularity. This framework was co-designed
with seventeen practitioners across two workshops and received directional support
in a follow-up survey of forty-three professionals (means $3.50$--$3.91$ on a
five-point scale), suggesting that the constructs address a recognized
practitioner need.

\subsection{Alignment with Evolving Tool Landscape}

Our results should be interpreted against a rapidly maturing product landscape in which reviewing LLM-generated code is becoming a distinct workflow: GitHub now publishes dedicated guidance for reviewing LLM-generated code~\cite{GitHubCopilotReviewAI} and reports that Copilot code review accounts for more than one in five code reviews on the platform~\cite{GitHubBlogCopilotReviews}. 

Several of the identified metaphors already have partial industrial counterparts. For example, variants of walk-through or overview support appear in CodeRabbit~\cite{CodeRabbitWalkthroughs} and in Harness AI~\cite{HarnessAIReview}. Our AI Judge metaphor also has clear analogs: Claude Code launches multiple reviewer agents that tag findings by severity~\cite{ClaudeCodeReview}, Codex focuses GitHub review comments on higher-priority P0/P1 issues~\cite{OpenAICodexGitHub}, and CodeAnt prioritizes findings using severity levels and risk signals such as blast radius and historical volatility~\cite{CodeAntBlogSeverity}. Graphite’s stacked PRs provide the closest current analogue to our chunking idea by splitting large tasks into independently reviewable units across multiple PRs rather than disentangling a single generated proposal. By contrast, token usage and sandboxing are usually treated as operational controls rather than reviewer-facing trust signals; for example, Claude Code documents token tracking for cost management and uses a default read-only, approval-based execution model. 

Overall, this suggests that the novelty of our proposal lies less in inventing isolated features than in combining them into a trust-calibrating, progressively disclosed workflow for reviewing multi-file LLM-generated changes.

\subsection{Hierarchical Review as Cognitive Scaffolding}
A further interpretation of the concepts produced in the design workshops is that developers were not only asking for more information, but for information organized across usable levels of abstraction. The proposed workflow moves from overview-level representations of the change, through file-level risk differentiation, to chunk-level review, while constructs such as zooming in/out, risk-per-file, risk-per-line, walk-through, and chunk make it possible to traverse these levels rather than inspect a large diff as an undifferentiated sequence of lines. This three-phase progression directly mirrors Shneiderman's visual information-seeking mantra: \enquote{overview first, zoom and filter, then details-on-demand}~\cite{shneiderman1996eyes}, and is consistent with cognitive load theory~\cite{sweller1988cognitive}: when a task contains many interacting elements, poorly organized representations consume processing capacity that could otherwise support understanding and decision-making, whereas chunking reorganizes complex, multi-file LLM suggestions into manageable semantic units that reduce extraneous load. It also aligns with working-memory and expertise research showing that people can actively maintain only a small number of independent chunks, while experts cope with complex domains by forming larger meaningful units rather than processing each element separately. In software specifically, this progressive disclosure supports top-down program comprehension~\cite{brooks1983towards}. Developers form high-level hypotheses about system behavior and subsequently validate them against granular code fragments, making a purely line-by-line diff view a poor fit for multi-file LLM-generated changes. By presenting architectural and data-flow diagrams during the overview phase, the tool equips developers with the context to generate these hypotheses, while localized snippet review provides the targeted details needed for verification. From this perspective, our constructs function as cognitive scaffolding for trust-calibration. The overview level supports initial situation awareness, file- and line-level risk signals guide attention allocation, and chunks provide reviewable units small enough for local judgment.

\subsection{Connection to Prior Work}

We position the contribution against three categories of prior work. First, the two-phase reading model of Gon\c{c}alves et al.~\cite{gonccalves2025code} is consistent with the workshop's three-level result; we extend it by inserting a file-level differentiation layer that becomes necessary when authorship is non-human. Second, the \emph{Code Review as Decision-Making} model~\cite{gullstrand_heander_code_2026} is adjacent prior work that motivates studying the analytical phase in isolation. CRDM's distinction between an orientation phase and an analytical phase is consistent with the observation that, when the reviewer is also the prompter, orientation collapses and the analytical phase becomes the locus of work. Third, the AI-as-reviewer paradigm represented by Vijayvergiya et al.~\cite{vijayvergiya2024ai} adopts an \emph{augmentation} stance in which the AI comments on human-authored code; the present work concerns the inverse configuration and is therefore complementary.

\subsection{Future Work}
\label{sec:future_work}

We identify several directions for future work, each driven directly by signals in the validation survey or by limitations of the current artifact.

\textbf{Free vs. Phase-Locked Navigation.} Five survey respondents recommended skipping the division into levels and letting the reviewer move freely between views. A within-subjects comparison of fixed-level and free-navigation variants of the prototype could disambiguate whether the three-level scaffolding aids attention allocation or constrains expert reviewers unnecessarily.

\textbf{Test-Coverage Integration.} Several respondents suggested integrating test-coverage data into the Judge view. This could strengthen the Judge construct as a trust signal by grounding its verdicts in observable test behavior, rather than in static analysis alone.

\textbf{Hands-On Longitudinal Prototype Evaluation.} The current evidence base is a survey of N=43 practitioners responding to a prototype walk-through. Evaluating a fully operational prototype by hands-on longitudinal use within development teams could upgrade the directional findings reported here to confirmatory evidence, and could surface potential adoption frictions. This could also confirm if the security cage is a viable construct for code review or better solved on the agent configuration level.

\section{Threats to Validity}
\label{sec:threats}

Following Lago et al.~\cite{lago2024threats}, we treat validity analysis as an active design concern rather than a post-hoc checklist. Because the study is a pragmatist participatory design study with a follow-up survey, we organize threats using Guba's qualitative trustworthiness framework~\cite{guba1981criteria}, since Guba's criteria align with the naturalistic, co-constructed character of the workshop data. For each criterion we name the threat, the mitigations operating in the design, and the residual risk that mitigation does not eliminate.

\textbf{Credibility.} Under the pragmatist stance, knowledge claims are evaluated by their utility in practice. Two mitigations operate in the study design. First, member checking was embedded in the workflow: Workshop~III opened by presenting the Workshop~II problem framings to the seven returning participants and inviting feedback before the co-design activity began (\Section{sec:method}). Second, the constructs and workflow are triangulated across elicitation modalities, appearing in Workshop~I post-its, Workshop~III sketches, and the validation survey free-text and Likert responses. Several threats nonetheless persist. The follow-up survey evaluated a semi-interactive prototype via a video demonstration rather than hands-on use in a realistic review task, so it tests \emph{perceived} usefulness rather than enacted usefulness. The survey respondents may carry acquiescence bias that inflate Likert ratings, and the survey is a single-snapshot measurement without a behavioral anchor. The label \emph{trust-calibration} is also an authorial synthesis: participants independently identified trust as the dominant challenge, and the term was mentioned at the start of Workshop III, but no participant challenged the label or the construct. The residual risk is that the constructs and workflow may be wanted and plausible to practitioners with relevant experience while not yet being shown to change reviewer behavior or outcomes in a live, multi-file review; a hands-on longitudinal evaluation is required to close this gap (see \Section{sec:future_work}).

\textbf{Transferability.} Under pragmatist epistemology, findings transfer by analytical generalization rather than statistical sampling, and the boundary conditions for transfer must be stated explicitly. Two boundaries apply. The participant profile defines the first: all workshop participants held professional software engineering roles, had used AI coding assistants for at least three months at least weekly, and had reviewed multi-file LLM-generated changes; claims should not be extended to junior developers, infrequent AI tool users, or teams where multi-file LLM-generated changes are not yet routine. The study period defines the second: the four workshops and the validation survey were conducted within a two-month window, which kept the AI tool environment effectively static across data collection (heavy use of Claude Code, ChatGPT, OpenAI Codex, GitHub Copilot, and Cursor; \Section{sec:method}). Forward-transfer is the more substantive temporal threat: as tool capabilities shift toward more agentic, multi-file, and self-explanatory behavior, the salience of individual constructs (\Section{sec:results}) may shift with them, with constructs such as \emph{solution plan} and \emph{walk-through} most exposed to absorption by the tools themselves. Recruitment also introduced volunteering bias: participants self-selected in response to a call, likely over-representing developers with positive prior experience with AI coding tools. Similar volunteering bias exists in the validation survey, since the survey was distributed through the same recruitment channels.

\textbf{Dependability.} Dependability concerns whether the synthesis process was repeatable and consistent across the research team. Four procedural mitigations operate in the design, each described in \Section{sec:method}: investigator triangulation through a fourth author who joined only at the convergent phases; pilot workshops approximately one week before each divergent workshop; consensus-based iterative joint coding across all four authors; and member checking of Workshop~II problem framings at the opening of Workshop~III. The replication package supports dependability by exposing the workshop protocols for inspection (\Section{sec:data_availability}). The residual risk is that investigator triangulation is partial rather than full: three of the four authors were continuous across phases.

\textbf{Confirmability.} Under the pragmatist stance, the researcher's situated judgment is a constitutive part of how meaning is produced; confirmability therefore concerns whether the resulting artifacts can be traced back to participant data rather than to author preference. Three traceability mitigations operate in the design. First, the Workshop~II prioritization of eleven problems was constrained by participant-assigned severity and feasibility ratings combined with cluster frequency, rather than by author preference alone (\Section{sec:method}). Second, the Workshop~III construct-admission rule required that constructs appear in multiple independent participant sketches and be corroborated by workshop discussion transcripts before being admitted, providing a stated rule that can be re-applied (\Section{sec:method}). Third, the validation survey filtering procedure for LLM-generated responses (54 of 97 responses removed on the basis of completion time, similarity to other responses, referrer link, and evident prompt leaking; \Section{sec:method}) is documented in the replication package together with the raw responses, so that the heuristic filter can be re-checked by external readers. The replication package thereby supports confirmability primarily, by exposing the audit trail from raw post-its and sketches to clusters and constructs (\Section{sec:data_availability}). The residual risk is that re-analysis remains possible but has not been performed, and that the AI-response filter is heuristic rather than validated.

\section{Conclusions}
\label{sec:conclusions}

Through a four-phase Double Diamond participatory design study with seventeen practitioners across two co-design workshops (seven of whom returned for the Develop phase) and a follow-up validation survey with forty-three software professionals, we identify \emph{trust-calibration} as the primary obstacle to effective review of LLM-generated multi-file changes (\RQ{1}). The participants converged on a three-level IDE workflow (overview, file-analysis, and code snippet review) supported by seven design constructs: chunk, risk-per-line, risk-per-file, judge, walk-through, zooming in/out, and security cage (\RQ{2}). A high-fidelity semi-interactive prototype operationalizing these constructs was rated above the neutral midpoint on every workflow level, and a majority of survey respondents expected reduced review effort (63\%) and reduced effort for trust assessment (52\%) compared with their existing tooling. The trust-calibration framing challenges the prevailing assumption that diff-view tooling scales to agent-authored changes; opacity of model confidence and the absence of chunk-level decomposition emerge as distinct, addressable failure modes. The constructs together form a conceptual framework that designers can use to reason about next-generation AI-ready code review tools.

\section{Data Availability}
\label{sec:data_availability}

Research data from this study, including workshop protocols for the Define and Develop workshops, the survey questionnaire, animations showing the prototype, source code for generation of figures and tables, and scanned and transcribed post-it notes and sketches from the workshops are available at \url{https://doi.org/10.5281/zenodo.20124352}. The replication package is licensed under MIT License and contains a README file with instructions for navigating the data. The high-fidelity semi-interactive prototype contains parts of proprietary software and cannot be shared in source code form.

\bibliographystyle{plainurl}
\bibliography{refs}

\end{document}